# The High Definition X-ray Imager (HDXI) Instrument on the *Lynx* X-Ray Surveyor

Abraham D. Falcone*[a], Ralph P. Kraft[b], Marshall W. Bautz[c], Jessica A. Gaskin[d], John A. Mulqueen[d], Doug A. Swartz[d], for the Lynx Science & Technology Definition Team

[a]Department of Astronomy & Astrophysics, Pennsylvania State University, 525 Davey Lab, University Park, PA 16802; [b]Harvard-Smithsonian Center for Astrophysics, 60 Garden Street, Cambridge, MA, 02138; [c]Massachusetts Institute of Technology, Kavli Institute for Astrophysics & Space Research, 77 Massachusetts Ave, Cambridge, MA, 02139; [d]NASA Marshall Space Flight Center, Huntsville, AL 35812

## ABSTRACT

The *Lynx* X-ray Surveyor Mission is one of 4 large missions being studied by NASA Science and Technology Definition Teams as mission concepts to be evaluated by the upcoming 2020 Decadal Survey. By utilizing optics that couple fine angular resolution (<0.5 arcsec HPD) with large effective area (~2 $m^2$ at 1 keV), *Lynx* would enable exploration within a unique scientific parameter space. One of the primary soft X-ray imaging instruments being baselined for this mission concept is the High Definition X-ray Imager, HDXI. This instrument would achieve fine angular resolution imaging over a wide field of view (~ 22 × 22 arcmin, or larger) by using a finely-pixelated silicon sensor array. Silicon sensors enable large-format/small-pixel devices, radiation tolerant designs, and high quantum efficiency across the entire soft X-ray bandpass. To fully exploit the large collecting area of *Lynx* (~30x Chandra), without X-ray event pile-up, the HDXI will be capable of much faster frame rates than current X-ray imagers. The planned requirements, capabilities, and development status of the HDXI will be described.

## 1. INTRODUCTION

The *Lynx* X-ray Surveyor is one of four large missions currently being studied by a NASA Science and Technology Definition Team (STDT) in order to provide input to the 2020 Decadal Survey process. *Lynx* is an X-ray observatory that will directly observe the dawn of supermassive black holes, reveal the invisible drivers of galaxy and structure formation, and trace the energetic side of stellar evolution and stellar ecosystems. It will provide huge improvements in X-ray sensitivity relative to existing and planned X-ray missions, including Athena, Chandra, and XMM. The current notional design for *Lynx* utilizes nested X-ray mirrors with a 3 meter diameter, which results in 2.3 $m^2$ effective area at 1 keV. The angular resolution of the mirror system is expected to be better than 0.5 arcsec on-axis and better than 1 arcsec within a 10 arcmin radius at the focal plane. The combination of this large effective area and fine angular resolution lead to a 50-100× improvement in throughput and sensitivity relative to existing and planned missions. The fine angular resolution throughout much of the field of view enables deep surveys with little to no source confusion, enabling e.g. the detection of very distant and dim AGN out to redshifts of z~10. These unprecedented characteristics, along with the excellent spectral resolution of *Lynx*, will enable: detailed studies of the first seed black holes, as well as their growth and evolution with time in the Universe, the impact of black hole system emission/feedback on multiple scales, the mapping of the intergalactic cosmic structures and the study of its impact on galactic growth and evolution, detailed studies of galaxy clusters, studies of stellar evolution and birth, along with the impacts of feedback on stellar systems, and the impact of X-ray stellar activity on extra-solar planet habitability. In addition to the above science topics, *Lynx* will be a major observatory class mission, which would enable myriad major-impact community driven science studies through a vibrant general observer program. For a more thorough overview of the *Lynx* mission and its science motivation, refer to Gaskin et al. [1].

*afalcone@astro.psu.edu

The current baseline *Lynx* design includes three instruments: 1 - The High Definition X-ray Imager (HDXI), 2 - The *Lynx* X-ray Microcalorimeter (LXM), and 3 - The X-ray Grating Spectrometer (XGS). The HDXI and the LXM are mounted on a translation table (see Figure 1), which allows one of them to be in the field of view at any given time. The grating array readout is mounted to the side, and is always in position for readout of the diffracted and offset X-rays, with its utility dependent on whether or not the grating itself is in position. The LXM is discussed by Bandler et al. [2]; the XGS is discussed by McEntaffer et al. [3] and Moritz et al. [4]; and in this paper, we will briefly describe the HDXI instrument.

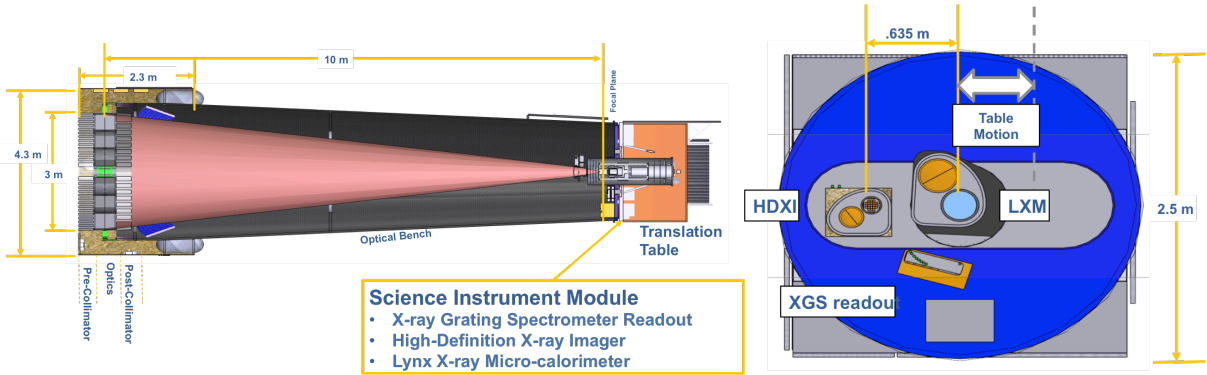

Figure 1. Initial/notional design of *Lynx* X-ray Surveyor telescope assembly (left) and focal plane layout (right).

## 2. HDXI OVERVIEW

The HDXI instrument is an imaging X-ray spectrometer that is designed to achieve a moderately wide field of view while simultaneously achieving fine angular resolution. In order to adequately oversample the < 0.5 arcsec point-spread-function of the *Lynx* mirrors, the HDXI is designed with pixels that span less than 0.33 arcsec per pixel, which corresponds to pixels with a pitch that is ≤16 μm. With a field of field-of-view requirement driven by the need to efficiently survey the X-ray sky for distant accreting active galactic nuclei, the notional focal plane instrument is designed to be 22×22 arcmin. The combination of these two requirements leads to a focal plane camera with ~16 megapixels that spans approximately 6.5 cm × 6.5 cm. The large format, combined with a need for moderate spectral resolution and rapid readout of the high X-ray throughput, leads to Silicon detectors as a natural choice for the HDXI focal plane instrument. Since the survey science also requires <1 arcsec resolution near the edges of the field-of-view (FOV), we are baselining multiple detectors to cover this FOV since this allows the detectors to be tilted in order to better match the curvature of the focal surface. The notional design includes 21 silicon detectors, with a layout as shown in Figure 2.

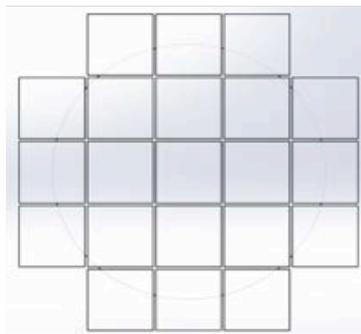

Figure 2. Schematic of detector layout. The circle shows the 22 arcmin diameter requirement on the FOV.

Table 1: Requirements and Parameters table for HDXI

| HDXI Parameter | Requirement | Science Drivers |
|---|---|---|
| Energy Range | 0.2 – 10 keV | Sensitivity to high-z sources |
| Field of view | 22 x 22 arcmin | Deep Survey efficiency<br>$R_{200}$ for nearby galaxies |
| Pixel size | 16 x 16 μm | Point source sensitivity<br>Resolve AGN from group emission |
| Read noise | ≤ 4 e- | Low-energy detection efficiency |
| Energy Resolution (FWHM) | 70 eV @ 0.3 keV<br>150 eV @ 5.9 keV | Low-energy detection efficiency; source and background spectra |
| Full-field count-rate capability | 8000 ct s-1 | No dead time for bright diffuse sources (e.g. Perseus Cluster or Cas A) |
| Frame Rate<br>Full-field<br>Window mode (20"x20") | > 100 frames s-1<br>> 10000 windows s-1 | Maximize low-energy throughput<br>Minimize background |
| Radiation Tolerance | TBD | maintain science capabilities for >10 years in orbit |

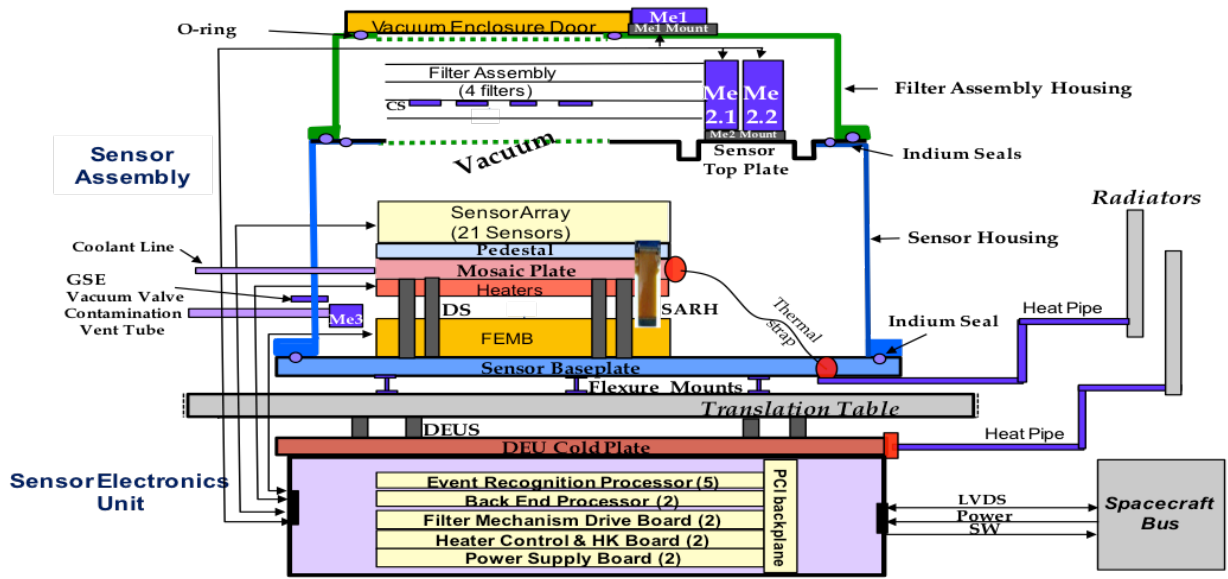

Figure 3. Schematic block diagram of HDXI from MSFC-ACO and GSFC-IDL combined studies.

The instrument requirements and parameters are driven by the science requirements derived by the *Lynx* STDT. These requirements and parameters are shown in Table 1. In addition to the FOV and pixel size requirements that were described above, the high throughput of the X-ray mirrors necessitates fast readout of the detectors in order to avoid pile-up of the individual X-rays that would otherwise lead to loss of energy resolution and flux measurement accuracy. The energy range is driven by the typical soft X-ray flux of the sources required to achieve the prime science goals of the mission.

Engineering and design studies have been carried out at the Marshall Space Flight Center Advanced Concept Office (MSFC - ACO) and at the Goddard Space Flight Center Instrument Design Lab (GSFC - IDL). These studies have led to significant refinement of the notional design for HDXI, as well as estimates of mass, power, volume, and cost. A notional instrument configuration that resulted from these studies is shown in Figures 3 and 4. This configuration includes the 21 sensor array that is mounted to a mosaic plate, which is cooled through a cold strap connection to the instrument chamber housing, which is passively cooled via external radiators. Thermal modeling indicated that this design was feasible for cooling the detectors, as well as an array of ASICs that were mounted on the front end mother board (FEMB). This sensor assembly vacuum chamber also contains an array of retractable filters: 2 for blocking UV/optical light, 1 protective blocking filter, and one with calibration X-ray sources. The electronics box contains boards for main processing, power, heater and filter control, housekeeping, and FPGA boards for grading and packaging X-ray event data. This initial design is intended as a reference point to estimate required resources and feasibility.

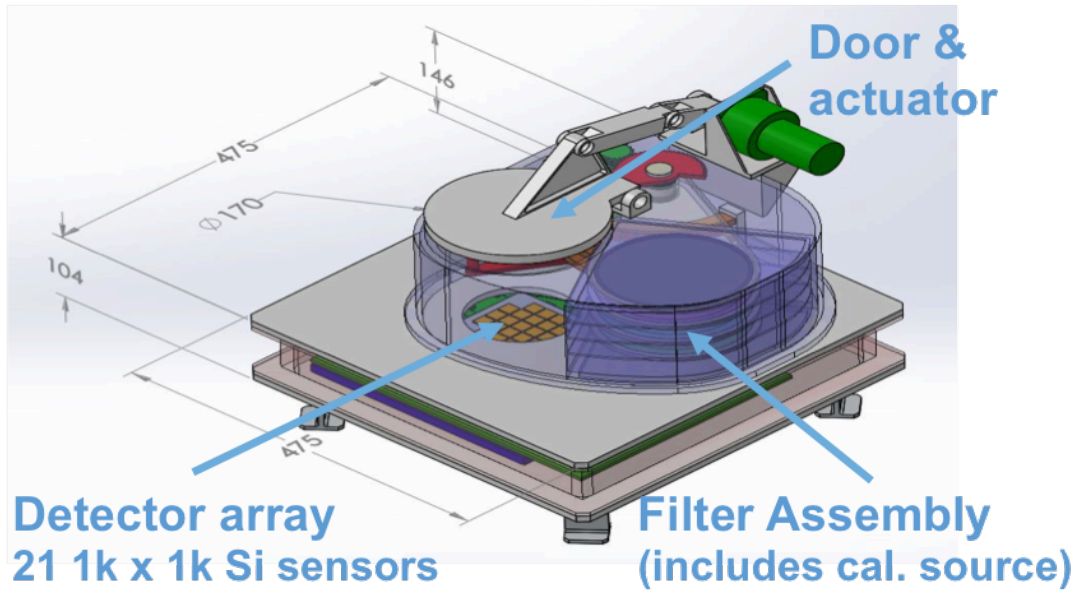

Figure 4. Detector assembly diagram for notional HDXI instrument, showing the filter assembly housing (designed at the GSFC - IDL) above the detector plane which resides inside a vacuum enclosure (shown with transparent walls).

## 3. HDXI DETECTORS

Three primary detector types are being considered for the notional HDXI instrument design. These detectors include two forms of active pixel silicon sensors, namely X-ray hybrid CMOS detectors (HCDs) and X-ray monolithic CMOS detectors (MCDs), as well as one form of CCD known as a digital CCD (see Figure 5). Each of the three sensor technologies has a clear path towards reaching TRL 5 by the time of mission adoption and phase B, which is projected to be in ~2024, but each of the technologies needs further research and development in order to reach this point.

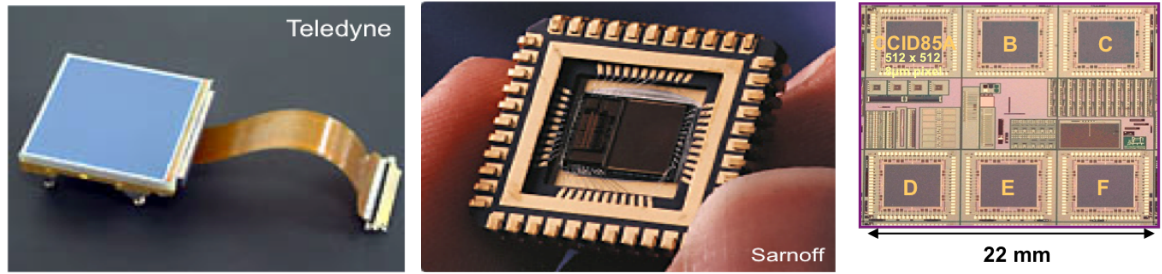

Figure 5. Three silicon detector technologies are currently being used to help specify HDXI parameters. They are: (1) hybrid CMOS active pixel sensors being developed by Teledyne and the Pennsylvania State University, (2) monolithic CMOS active pixel sensors being developed by Sarnoff and Harvard-Smithsonian Astrophysical Observatory, and (3) digital CCDs being developed by Lincoln Labs and the Massachusetts Institute of Technology.

### 3.1 X-ray Hybrid CMOS Detectors

X-ray hybrid CMOS detectors (HCDs) are active pixel sensors that are made by bonding a silicon detection layer, with an optimized thickness and resistivity, to another silicon layer containing readout-integrated circuitry (ROIC) in each pixel. They are back-illuminated devices offering high quantum efficiency over the full soft X-ray band pass (0.2-10 keV) with fully depleted silicon and deep depletion depths (>100 micron). These X-ray detectors are being developed in a collaborative program between Teledyne Imaging Systems and the Pennsylvania State University. These devices are inherently radiation hard due to the fact that the charge is not transferred across the device, and is instead, readout directly from each pixel. They offer low power relative to existing CCDs, and they have displayed high readout rates with multiplexed readout lines (>5 MHz per line through 32 lines). X-ray Hybrid CMOS detectors that were developed for other purposes have been demonstrated to TRL 9 in a recent rocket flight (Chattopadhyay et al. 2018) and non-X-ray versions have flown on recent observatories (e.g. OCO-2), but these devices have moderate noise levels and large pixels. Newer test devices have been fabricated and tested with lower noise (~5.5 e- RMS), small pixels (12.5 micron pitch), and on-chip correlated double sampling (Hull et al. 2018). Some of these devices are also capable of event driven readout and digitization on-chip. Future developments aim to achieve low readnoise (<4 e-) while retaining the already-achieved small pixels with on-chip correlated double sampling (CDS) and digitization, deep depletion, high speed, low power, and radiation hardness.

### 3.2 Monolithic CMOS Detectors

Monolithic CMOS sensors are under development by Smithsonian Astrophysical Observatory and Sarnoff Research Institute for a wide variety of X-ray astrophysics applications, including *Lynx*. Low noise (~2.9 e- rms), low dark current, and good sensitivity to low energy X-rays (B $K_\alpha$ and C $K_\alpha$) were demonstrated on back-illuminated 1k x 1k 16 μm pixel pitch sensors built on 10 μm thick epitaxial silicon (Kenter *et al*. 2017). Intrinsic advantages of this technology include radiation hardness, high frame rates, low power, and high levels of on-chip integration. The next generation of sensors under development will be built on p-pixel technology with n-type Si substrates (Kenter et al. 2018). Collection of holes instead of electrons will reduce the read noise and dark current, and increase the radiation hardness. PMOS technology potentially offers several other advantages for X-ray astronomy over the convention NMOS technology. The effect of trapping is greatly reduced at the sensor surface, thus potentially improving the energy resolution below 1 keV for back-illuminated devices. This technology will also virtually eliminate the random telegraph signal. Future developments aim to achieve deep depletion with the addition of a bias layer to the back surface to increase the quantum efficiency to 10 keV.

### 3.3 Digital CCDs

The Digital Charge Coupled Device (CCD) architecture under development for *Lynx* and other programs at Massachusetts Institute of Technology (MIT) Lincoln Laboratory features a CCD detector with multiple high-speed (up to 5 MHz) outputs that are compatible with fast, low-power CMOS drive and signal processing circuitry. Two parallel development paths focus on the detector and the signal processing circuitry, respectively. First generation digital CCDs achieve noise of 5.5 electrons (RMS) at 2.5 MHz pixel rates and excellent charge transfer efficiency with ± 1.5 V (CMOS compatible) charge transfer clocks (Bautz et al. 2018). NASA has funded a strategic astrophysics technology program to further develop this technology over the coming year. Future developments aim to achieve lower-noise, radiation-tolerant versions of these devices.

## 4. SUMMARY

Each of the above detector technologies is capable of achieving some of the requirements of HDXI, but none of these detector technologies can currently achieve all of the requirements simultaneously. Technology development over the next ~6 years will bring at least one of these technologies to TRL 5 for the start of the *Lynx* mission. While other detectors may be capable of achieving the requirements in the future and while further trade studies will be carried out to choose the optimum detector, these three options have enabled the notional instrument to be specified in a way that has proven feasibility. The notional HDXI design has shown that it can accommodate any of these three detector types and that reasonable mass, power, and cost are achievable. Through multiple MSFC-ACO and GSFC-IDL runs, we have developed a notional design for the HDXI instrument, and we have shown that it can be accommodated within reasonable constraints while fulfilling requirements that satisfy the groundbreaking science drivers identified by the *Lynx* Science and Technology Definition Team.